\documentclass[aps, amsfonts,amsmath,amssymb,nofootinbib,twocolumn,superscriptaddress]{revtex4}

\usepackage{graphics}
\usepackage{hyperref}
\usepackage{epsfig}
\usepackage{color}
\usepackage{multirow}
\usepackage{bm}

\begin{document}

\title{
Field evolution of magnons in $\alpha$-RuCl$_3$ by high-resolution polarized terahertz spectroscopy
}

\author{Liang Wu}
\email{liangwu@sas.upenn.edu}
\affiliation{Department of Physics, University of California, Berkeley CA 94720, USA}
\affiliation{Materials Science Division, Lawrence Berkeley National Laboratory, Berkeley CA 94720, USA}
\affiliation{Department of Physics and Astronomy, University of Pennsylvania, Philadelphia, Pennsylvania 19104, USA}

\author{A. Little}
\affiliation{Department of Physics, University of California, Berkeley CA 94720, USA}
\affiliation{Materials Science Division, Lawrence Berkeley National Laboratory, Berkeley CA 94720, USA}

\author{E. E. Aldape}
\affiliation{Department of Physics, University of California, Berkeley CA 94720, USA}

\author{D. Rees}
\affiliation{Department of Physics, University of California, Berkeley CA 94720, USA}
\affiliation{Materials Science Division, Lawrence Berkeley National Laboratory, Berkeley CA 94720, USA}

\author{E. Thewalt}
\affiliation{Department of Physics, University of California, Berkeley CA 94720, USA}
\affiliation{Materials Science Division, Lawrence Berkeley National Laboratory, Berkeley CA 94720, USA}

\author{P. Lampen-Kelley}
\affiliation{Department of Materials Science and Engineering, University of Tennessee, Knoxville, TN 37996, U.S.A.}
\affiliation{Materials Science and Technology Division, Oak Ridge National Laboratory, Oak Ridge, TN, 37830, U.S.A.}

\author{A. Banerjee}
\affiliation{Neutron Scattering Division, Oak Ridge National Laboratory, Oak Ridge, Tennessee 37830, USA.}

\author{C. A. Bridges}
\affiliation{Chemical Sciences Division, Oak Ridge National Laboratory, Oak Ridge, Tennessee 37830, USA.}

\author{J.-Q. Yan}
\affiliation{Materials Science and Technology Division, Oak Ridge National Laboratory, Oak Ridge, TN, 37830, U.S.A.}

\author{D. Boone}
\affiliation{Department of Applied Physics, Stanford University, Stanford, California 94305, USA}
\affiliation{Stanford Institute for Materials and Energy Sciences, SLAC National Accelerator Laboratory, Menlo Park, California 94025, USA}

\author{S. Patankar}
\affiliation{Department of Physics, University of California, Berkeley CA 94720, USA}
\affiliation{Materials Science Division, Lawrence Berkeley National Laboratory, Berkeley CA 94720, USA}

\author{D. Goldhaber-Gordon}
\affiliation{Department of Physics, Stanford University, Stanford, California 94305, USA}
\affiliation{Stanford Institute for Materials and Energy Sciences, SLAC National Accelerator Laboratory, Menlo Park, California 94025, USA}

\author{D. Mandrus}
\affiliation{Department of Materials Science and Engineering, University of Tennessee, Knoxville, TN 37996, U.S.A.}
\affiliation{Materials Science and Technology Division, Oak Ridge National Laboratory, Oak Ridge, TN, 37830, U.S.A.}

\author{S. E. Nagler}
\affiliation{Neutron Scattering Division, Oak Ridge National Laboratory, Oak Ridge, Tennessee 37830, USA.}

\author{E. Altman}
\affiliation{Department of Physics, University of California, Berkeley CA 94720, USA}

\author{J. Orenstein}
\email{jworenstein@lbl.gov}
\affiliation{Department of Physics, University of California, Berkeley CA 94720, USA}
\affiliation{Materials Science Division, Lawrence Berkeley National Laboratory, Berkeley CA 94720, USA}

\date{\today}

\begin{abstract}

The Kitaev quantum spin liquid (KSL) is a theoretically predicted state of matter whose fractionalized quasiparticles are distinct from bosonic magnons, the fundamental excitation in ordered magnets. The layered honeycomb antiferromagnet $\alpha$-RuCl$_3$ is a KSL candidate material, as it can be driven to a magnetically disordered phase by application of an in-plane magnetic field, with $H_c \sim 7$ T. Here we report a detailed characterization of the magnetic excitation spectrum of this material by high-resolution time-domain terahertz (THz) spectroscopy. We observe two sharp magnon resonances whose frequencies and amplitudes exhibit a discontinuity as a function of applied magnetic field, as well as two broader peaks at higher energy. Below the N\'eel temperature, we find that linear spin wave theory can account for all of these essential features of the spectra when a $C_3$-breaking distortion of the honeycomb lattice and the presence of structural domains are taken into account.

\end{abstract}

\maketitle

\section{Introduction}
The quantum spin liquid (QSL) is an exotic phase of matter characterized by a disordered yet highly entangled ground state. Geometrically frustrated magnets with, for example, a triangular arrangement of spins have been predicted to host such states. Another promising route to a QSL is the Kitaev honeycomb, which consists of spin-1/2 particles arranged on a honeycomb lattice. \cite{kitaev2003fault, kitaev2006anyons} In this model, anisotropic Ising-like exchange interactions between nearest neighbors give rise to frustration.
The ground state is a gapless Z$_2$ spin liquid, with excitations taking the form of itinerant Majorana quasiparticles and static fluxes.

\par The Kitaev honeycomb is of recent experimental interest, as the anisotropic interactions characteristic of the model can manifest in real materials,~\cite{jackeli2009mott, chaloupka2010kitaev} in particular transition metal compounds with strong spin-orbit coupling (SOC) such as the Na and Li iridates~\cite{williams2016incommensurate, biffin2014noncoplanar,modic2014realization, takayama2015hyperhoneycomb} and $\alpha$-RuCl$_{3}$. \cite{kim2015kitaev, plumb2014alpha} Despite the presence of a Kitaev term in the effective spin Hamiltonian, these materials order magnetically at low temperatures~\cite{liu2011long, choi2012spin, chun2015direct, biffin2014noncoplanar, takayama2015hyperhoneycomb, sears2015magnetic} indicating that they host interactions beyond Kitaev exchange. Characterizing these interactions can help to navigate the rich phase diagrams of these materials, wherein one may approach a quantum-disordered state by applying external perturbations such as fields or chemical substitution~\cite{yadav2016kitaev}.

\par $\alpha$-RuCl$_{3}$ has risen to prominence as a candidate Kitaev system, driven by the availability of single crystals suitable for inelastic neutron scattering (INS)~\cite{banerjee2016proximate,ran2017spin} and optical spectroscopy \cite{little2017antiferromagnetic, ponomaryov2017unconventional, wang2017magnetic,  shi2018field}, as well as the observation that magnetic order disappears in an in-plane magnetic field $H_c\sim$ 7.5 T~\cite{sears2017phase, baek2017observation, zheng2017gapless,leahy2017anomalous, banerjee2018excitations}. In this material, quasi-2D layers of Ru$^{3+}$ atoms surrounded by Cl$_6$ octahedra are arranged on a honeycomb lattice. The combination of octahedral crystal field splitting, electron correlations, and SOC gives rise to a Mott-insulating state with a localized J$_{eff}$ = 1/2 moment on each Ru$^{3+}$ site~\cite{kim2015kitaev}. The quasi-2D layers are stacked and van der Waals coupled to form bulk $\alpha$-RuCl$_{3}$.  Such layered magnets are of particular interest because they can be assembled and stacked with other 2D materials, forming heterostructures with potentially topological phases~\cite{soumyanarayanan2016emergent}.

\par Spectroscopic probes such as INS~\cite{banerjee2016neutron, banerjee2016proximate,banerjee2018excitations}, Raman scattering~\cite{sandilands2015scattering}, and THz absorption~\cite{little2017antiferromagnetic, wang2017magnetic, ponomaryov2017unconventional, shi2018field} have been employed to characterize magnetic fluctuations in $\alpha$-RuCl$_{3}$ and test for the existence of, or proximity to, a QSL phase. Below T$_{N}$ = 7 K, the ground state has zigzag antiferromagnetic order~\cite{sears2015magnetic}, as shown in Fig. 1. (b). In the ordered phase and in zero applied magnetic field, INS measurements observed peaks consistent with magnons together with a continuum of scattering centered at $\bf{Q}$ = 0 $(\Gamma$-point) that was seen as well by Raman spectroscopy. This continuum was found to persist at fields above $H_c$, as well as at temperatures above $T_{N}$ at zero field, and was interpreted as a possible signature of fractionalized excitations, \textit{i.e.}, Majorana fermions and Z$_2$ vortices. However, it has also been suggested that the continuum reflects the breakdown of coherent magnons originating from strong anharmonicity in the magnon Hamiltonian.~\cite{winter2017breakdown} THz absorption measurements~\cite{little2017antiferromagnetic} showed that below T$_{N}$ the majority of the $\Gamma$-point spectral weight at low energies was accounted for by spin waves, and furthermore, that the contribution from a magnetic continuum did not grow with increasing magnetic field, up to $H_c$.
		
Recent measurements have explored in detail the region of the phase diagram near the critical field for the loss of magnetic order. Thermodynamic and transport measurements, including specific heat~\cite{kubota2015successive, wolter2017field, sears2017phase}, nuclear magnetic resonance~\cite{baek2017observation}, and thermal transport\cite{leahy2017anomalous, hirobe2017magnetic, hentrich2018unusual, yu2018ultralow} indicate a transition to a gapped magnetically disordered state.  However, varying interpretations of the nature of that state and its low-energy excitations leave the question of a transition to a QSL at or near $H_c$ unresolved. Recent experiments reporting a quantized thermal Hall effect \cite{kasahara2018majorana} for off-axis applied fields, a signature of chiral Majorana modes, suggest that a topological phase may exist in the vicinity of $H_c$.

Measurements of the magnetic excitation spectrum using time-domain THz spectroscopy (TDTS) can aid theoretical understanding of the $\alpha$-RuCl$_3$ phase diagram by constraining the effective spin Hamiltonian parameters. TDTS probes this spectrum with high sensitivity and energy resolution, yielding an absolute measurement of the imaginary part of the dynamic magnetic susceptibility at zero wavevector, $\chi^{\prime\prime}(\omega, Q = 0)$ in the energy range 0.1 to 1.7 THz, or 0.4 to 7.0 meV ~\cite{little2017antiferromagnetic}. In Section II we describe THz absorption measurements that fully characterize $\chi^{\prime\prime}(\omega, Q = 0)$ associated with the antiferromagnetic state of $\alpha$-RuCl$_3$ as a function of static field $\bf{H}$ and THz probe field $\bf{B_{THz}}$. We observe four resonances whose frequency and amplitude exhibit a complex dependence on applied field that depends strongly on the relative orientation of $\bf{H}$ and $\bf{B_{THz}}$. We use the absolute determination of $\chi^{\prime\prime}(\omega, Q = 0)$ provided by THz absorption to track the the dependence of the spin wave spectral weight on $H$ for $\bf{B_{THz}}\ \|\ \bf{H}$ and $\bf{B_{THz}}\perp\bf{H}$. These spectral weights are then compared with the static susceptibility, $\chi(\omega = 0, Q = 0)$ to determine the relative contributions of spin wave vs. continuum to the total weight of magnetic fluctuations at zero wave vector. In Section III we compare our experimental results with calculations based on linear spin wave theory (LSWT). Surprisingly, we find that LSWT can account for all the essential features of the spectra -- the number of modes, their spectral weight and optical selection rules, the variation of resonant frequency with $H$, and a discontinuity in mode frequency and amplitude at a low field of $\sim$ 1.5 T. Achieving this description requires considering a $C_3$-breaking distortion of the honeycomb lattice and the resulting multi-domain structure, as well as a refinement of existing parameterizations of the effective spin Hamiltonian to account for a zero-field splitting of the lowest frequency spin waves. In addition, the contribution to the spectrum from two-magnon states is clearly identified. Finally, Section IV summarizes the conclusions of our study. Finally, Section IV summarizes the conclusions of our study.  \newline

\section{Experimental Results}
\subsection{Definition of axes}
To guide the polarized TDTS measurements, the optical anisotropy of $\alpha$-RuCl$_3$ samples was first characterized by measuring their transmitted THz electric field amplitude when rotated between crossed linear polarizers. Fig. 1(a) shows a typical room temperature scan of transmission as a function of angle of rotation about the optic axis. The nearly four-fold pattern, observed in all samples studied, indicates breaking of $C_3$ symmetry. This result is consistent with X-ray diffraction measurements that indicate a $\sim 0.2 \%$ elongation of one of the Ru-Ru bonds and a monoclinic $C2/m$ space group~\cite{johnson2015monoclinic, cao2016low}.  Fig. 1(b) depicts a Ru honeycomb layer that forms this structure, where $x$, $y$, and $z$ label the Ising axis of the Kitaev exchange term on the Ru-Ru bonds. An elongation in the direction of one the bonds (the one labeled by $z$ in the sketch) defines the $\bf{b}$ axis of the monoclinic structure. The color of the atoms illustrates the zigzag antiferromagnetic order that arises below the N\'eel temperature (T$_N$).

\begin{figure}[htp]
\centering
\includegraphics[width=1.0\columnwidth]{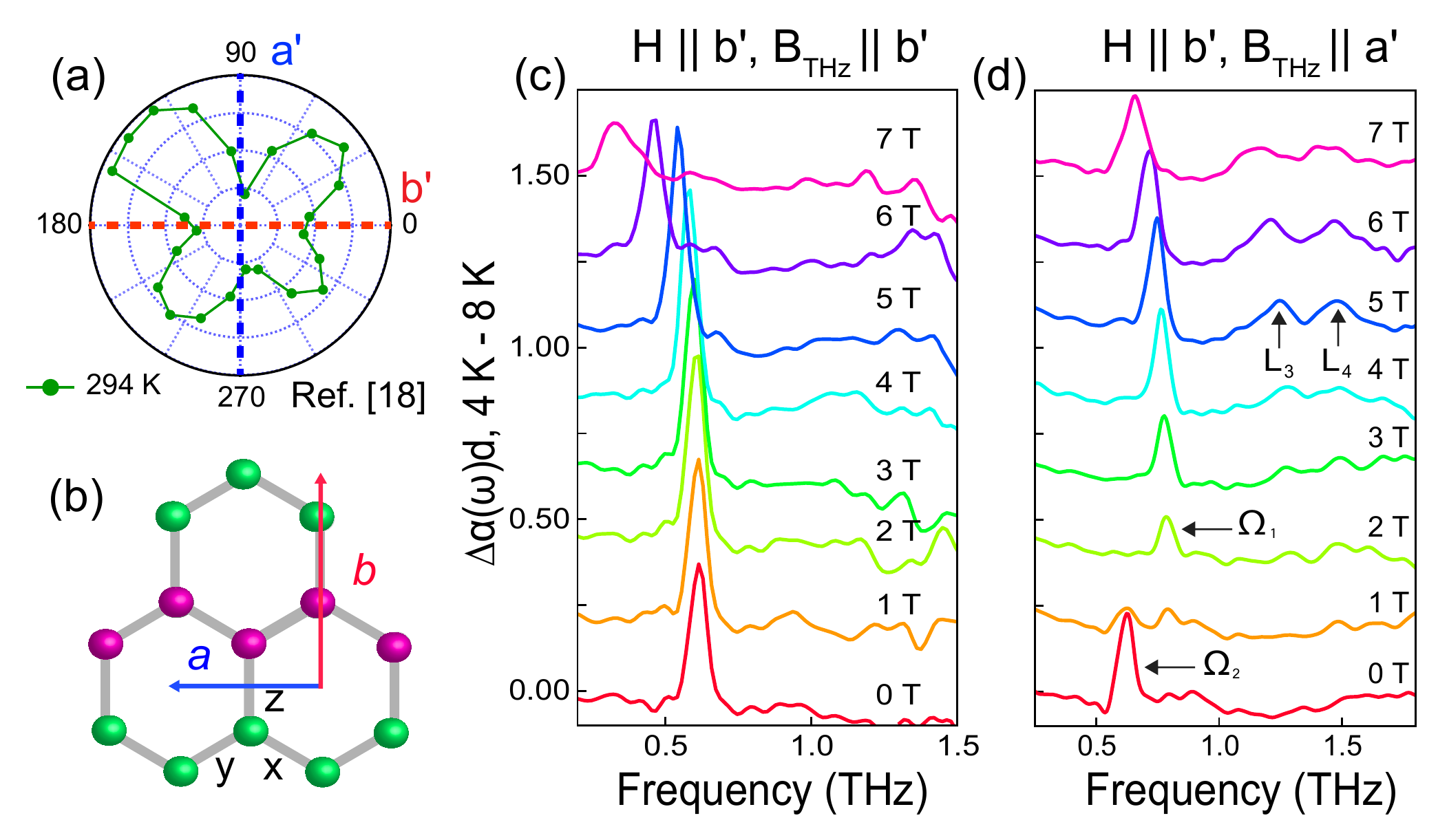}
\caption{(a) Transmitted THz electric field amplitude at T = 294 K as a function of sample angle. Blue and red lines represent the minimum transmission axes at $\bf{a^\prime}$ and $\bf{b^\prime}$ (b)  Schematic of honeycomb structure showing ${\bf a}$ and $\bf{b}$ monoclinic axes relative to Ru-Ru bonds. Color of atoms illustrates zigzag order. Bond labels x, y, and z denote the component of the spin interacting along a given bond in the Kitaev model. (c) Magnon absorption as a function of frequency for $\bf{H\ \|\ b^\prime\ \| \ B_{THz}}$ and $\bf{H\ \|\ b^\prime \perp B_{THz}}$ respectively. The magnon contribution is extracted from the total THz absorption by subtracting a reference at T = 8 K, above T$_{N}$, from a T = 4 K spectrum at each field. Traces are offset for clarity.
}
\label{fig1}
\end{figure}

\par The absence of nodes in the polar pattern in Fig. 1(a) indicates that the local $C2/m$ symmetry is broken globally by the presence domains of the three equivalent orientations of monoclinic distortion, which are rotated 120$^\circ$ from one another. A single domain $C2/m$ crystal would exhibit zero transmission for THz fields polarized parallel to the $\bf{a}$ or $\bf{b}$ axes, which is not seen in any of the samples we have studied. On the other hand, in a sample containing equal populations of three domains the optical anisotropy of each would be effectively canceled and the THz transmission between crossed polarizers would vanish for all angles. What we observe instead is the intermediate case, where unequal domain population gives rise to weak residual anisotropy. To confirm the presence of multiple domains, we performed scanning X-ray micro-Laue diffraction measurements~\cite{tamura2003scanning} that indeed revealed the presence of all three domains with spatially varying populations as discussed Appendix A, section 3. This multi-domain character, as we will show, is essential to understanding the THz absorption spectra in the zigzag state as a function of magnetic field.

Because of the low effective symmetry, the directions of minimum transmission in Fig. 1(a) do not coincide with the monoclinic axes of a single domain, although they will be close to those of a dominant domain. In this study, we reference our THz polarization and external magnetic field $\bf{H}$ to the two directions of minimum transmission, which we label as $\bf{a^\prime}$ and $\bf{b^\prime}$ to distinguish them from the monoclinic axes of a single domain. We measure the absorption coefficient $\alpha(\omega)$ with the THz probe field in the honeycomb plane, $\bf{B_{THz}}$, oriented parallel to $\bf{a^\prime}$ and $\bf{b^\prime}$, and in both cases we compare measurements with in-plane $\bf{H}$ parallel and perpendicular to $\bf{B_{THz}}$.

\subsection{Magneto-optical THz spectroscopy} The magnetic dipole contribution to $\alpha(\omega)$ that is associated with the presence of antiferromagnetic order can be isolated by subtracting spectra measured at T = 8 K, which is sufficiently above T$_N$ such that magnons are no longer present, from spectra in the ordered phase at T = 4 K (see Appendix A, section 5). The residual spectrum omits any magnetic contribution that does not change when crossing T$_N$. Figs. 1(c) and (d) show differential (4 K - 8 K) absorption spectra, $\Delta\alpha(\omega) d$, for a sample of thickness $d\sim1$ mm for $\textbf{H}$ parallel to $\bf{b^\prime}$. In the parallel ($\bf{B_{THz}}\ \|\ \textbf{H}$) channel (Fig. 1(c)), a single magnon is observed at $\Omega_1$ = 2.6 meV (0.62 THz) for $\bf{H}$ = 0, which shifts to lower energy and broadens as the field is increased. The spectra measured with $\bf{B_{THz}}\perp \textbf{H}$, shown in Fig. 1(d), are more complex in that the frequency and spectral weight appear to vary non-monotonically in field. In addition, two broader features, which we denote by $L_3$, and $L_4$, appear in the energy range 4-6 meV and become more strongly absorbing as the field is increased.

\begin{figure*}[htp]
\centering
\includegraphics[width=1.8\columnwidth]{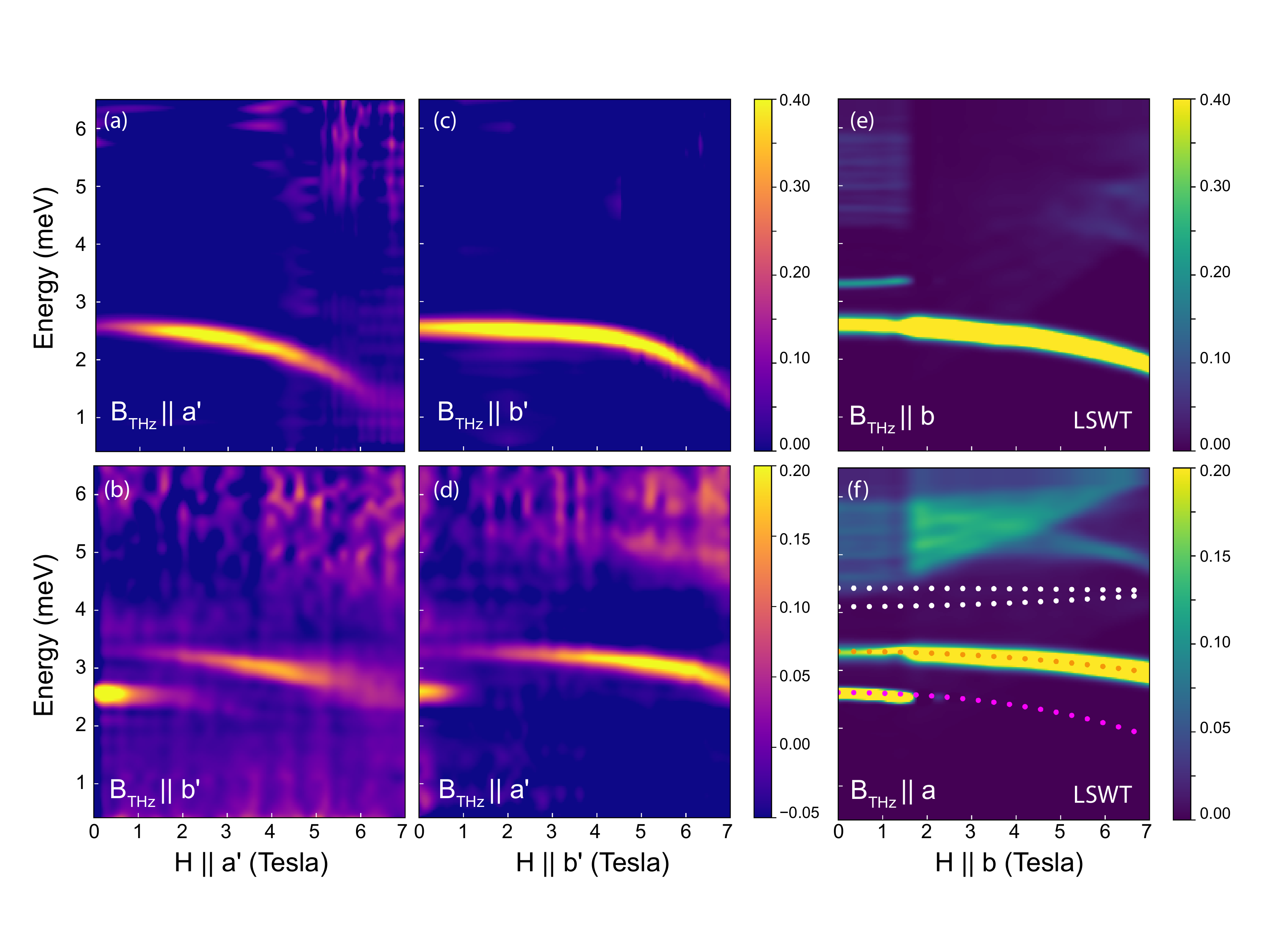}
\caption{ Magnon energies and absorption strengths at $\bf{Q}$ = 0 as a function of external in-plane magnetic field, H. Experimental data is in panels (a)-(d). Magnon absorption was extracted by subtracting the 8 K spectra from the 4 K spectra at each value of H. Spectra were taken in 0.2 T steps from 0 - 5 T and in 0.1 T steps from 5 - 7 T; intermediate field values are interpolated. The mode dispersion is shown for four configurations of H and the THz probe field, $\bf{B_{THz}}$ relative to $\bf a^\prime$ and $\bf b^\prime$: (a) and (c) show $\bf{H}\ \|\ \bf{B_{THz}}$ along the $\bf a^\prime$ and $\bf b^\prime$ directions respectively, while (b) and (d) show $\bf{H}\perp\bf{B_{THz}}$. Note the difference of color scales: absorption in the parallel configuration is roughly twice as strong. Panels (e) and (f) show LSWT calculations for absorption in $\bf{H}\ \|\ \bf{b}$ with the probe field parallel and perpendicular, respectively. Solid dots overlaid on (f) represent mode energies predicted by LSWT. The orange and pink dots coincide with observed $\Omega_1$ and $\Omega_2$. Two higher energy modes (white dots) are forbidden by selection rules and do not contribute to THz absorption. Intensity in the region 4 - 6 meV, consistent with observed $L_3$ and $L_4$, results from 2-magnon absorption.
}

\label{fig2}
\end{figure*}

\par The evolution of the spectra with $H$ is greatly clarified by the color scale plots in Fig. 2, which illustrate the magnitude of $\Delta\alpha d$ in the $\hbar\omega-H$ plane. Panels (a), and (b) show spectra with $\bf{H}\ \|\ \bf{a^\prime}$, for $\bf{B_{THz}}$ parallel and perpendicular to $\bf{H}$, respectively. Panels (c), and (d) are the corresponding spectra for the $\bf{H}\ \|\ \bf{b^\prime}$ configuration. Panels (e) and (f) show fits obtained by LSWT calculations discussed below.

We first note that the anisotropy with respect to rotation of the crystal by 90$^\circ$ is weak, that is, the pair of panels (a) and (b) share the same qualitative features as panels (c) and (d), with overall amplitude difference of only $\sim 2$. As we discuss below, LSWT predicts a much larger anisotropy in the dynamic susceptibility between the two principal ($\bf{a}$ and $\bf{b}$) axes of a single zigzag domain. We interpret the observed weak anisotropy to be further evidence for the presence of multiple domains. The width of the peaks remains relatively constant until around H $\sim$ 5 T at which point they start to broaden~\cite{little2017antiferromagnetic}. The broadening occurs more rapidly for $\bf{H\ \|\ a^{\prime}}$ as is apparent in Fig. 2 (a), where the $\Omega_1$ magnon becomes diffuse approaching 7 T. This is an indication that for $\bf{H\ \|\ a^{\prime}}$ the system is close to the critical point and corrections to the spin wave expansion become relevant (see Appendix B, section 3).

\par A far stronger contrast is seen when comparing spectra with $\bf{B_{THz}}\ \| \ \bf{H}$ (panels (a) and (c)) to $\bf{B_{THz}}\perp\bf{H}$ (panels (b) and (d)). For  $\bf{B}\ \|\ \bf{H}$ we observe a single mode that shifts to lower frequency with increasing $H$, with the field-induced mode softening slightly more pronounced with $\bf{H}\ \|\ \bf{a^\prime}$. For $\bf{B_{THz}} \perp \bf{H}$ the color plots show clearly that, rather than a single mode with a non-monotonic dependence of energy on field, there are in fact two distinct low energy modes. At $H = 0$ there is a strong mode, $\Omega_1$ = 2.6 meV, and a much weaker one, $\Omega_2$ = 3.3 meV. We note in particular the 0.7 meV splitting between these modes, which informs our LSWT calculations. As ${H}$ increases the spectral weight of $\Omega_1$ decreases rapidly and then shifts to $\Omega_2$ for H $\sim 1.5$ T. Surprisingly, the total spectral weight at this crossover field is close to zero.

\par The absorption features centered at $L_3 = 5.2$ meV and $L_4 = 6.2$ meV at H = 4 T, grow with increasing $H$ and persist as $H$ approaches $H_c$. An exact diagonalization study of $\alpha$-RuCl$_3$ associated eigenstates in this energy range with a two-magnon continuum~\cite{winter2018probing}. Our results for $\chi^{\prime\prime}(\omega)$ using LSWT described in the next section [and shown in Fig. 4(f)] account for the field and polarization dependence of $L_3$ and $L_4$, and confirm their origin as two-magnon excitations in the longitudinal response, that is, $\bf{B_{THz}}$ parallel to the zigzag wavevector.

\subsection{Magnetic susceptibility}
The differential THz absorption is directly related to the imaginary part of the zero wave vector dynamic susceptibility $\chi^{\prime\prime}(\omega$), that is,
\begin{equation}
\Delta\alpha(\omega)\cong\frac{n}{2}\frac{\omega}{c}\chi^{\prime\prime}(\omega),
\end{equation}
where $c/n$ is the speed of light in $\alpha$-RuCl$_3$ in the THz regime, which is determined independently (see Appendix A, section 2). The thermodynamic sum rule, derived from the Kramers-Kronig relation, relates $\chi^{\prime\prime}(\omega)$ to the dc magnetic susceptibility, $\chi(0)$. With this sum rule, the contribution to $\chi(0)$ from $\bf{Q}$ = 0 spin waves can be determined from the spectral weight of the spin wave peaks,
\begin{equation}
\chi_{sw}(0)\equiv\frac{2}{\pi}\int_0^\infty\frac{\chi^{\prime\prime}_{sw}(\omega)}{\omega}d\omega,
\end{equation}
where the subscript $sw$ denotes the component of susceptibility originating from spin wave resonances. By comparing $\chi_{sw}(0)$ to $\chi(0)$ we can place an upper bound on the spectral weight not accounted for by spin waves, i.e., a magnetic continuum~\cite{little2017antiferromagnetic}.

We evaluate $\chi^{\prime\prime}_{sw}(\omega)$ by fitting a Lorentzian function to the THz resonances (see Appendix A, section 6). The resulting $\chi_{sw}(0)$ is plotted in Fig. 3, for each of the four configurations of $\bf{H}$ and $\bf{B_{THz}}$ shown in Fig. 2. Also shown in Fig. 3 is $\chi_{\|}(0)$ as a function of magnetic field, which is defined by the change in magnetization resulting from a $\bf{\delta}H$ parallel to $\bf{H}$. Note that $\chi_{sw}(0)$ in the $\bf{B_{THz}}\|\bf{H}$ channel tracks $\chi_{\|}(0)$ as they both increase with increasing field. The difference $\chi_{\|}(0)-\chi_{sw}(0)$, which is an upper bound on the spectral weight of a magnetic continuum, persists but does not increase until $H$ becomes close to $H_c$. Finally, we note a small feature near 5.5 T in the parallel configuration for both the $\bf{a^\prime}$ and $\bf{b^\prime}$ curves, roughly consistent with a proposed intermediate phase in the $5-7$ T range~\cite{banerjee2018excitations}.

\begin{figure}[htp]
\centering
\includegraphics[width=0.85\columnwidth]{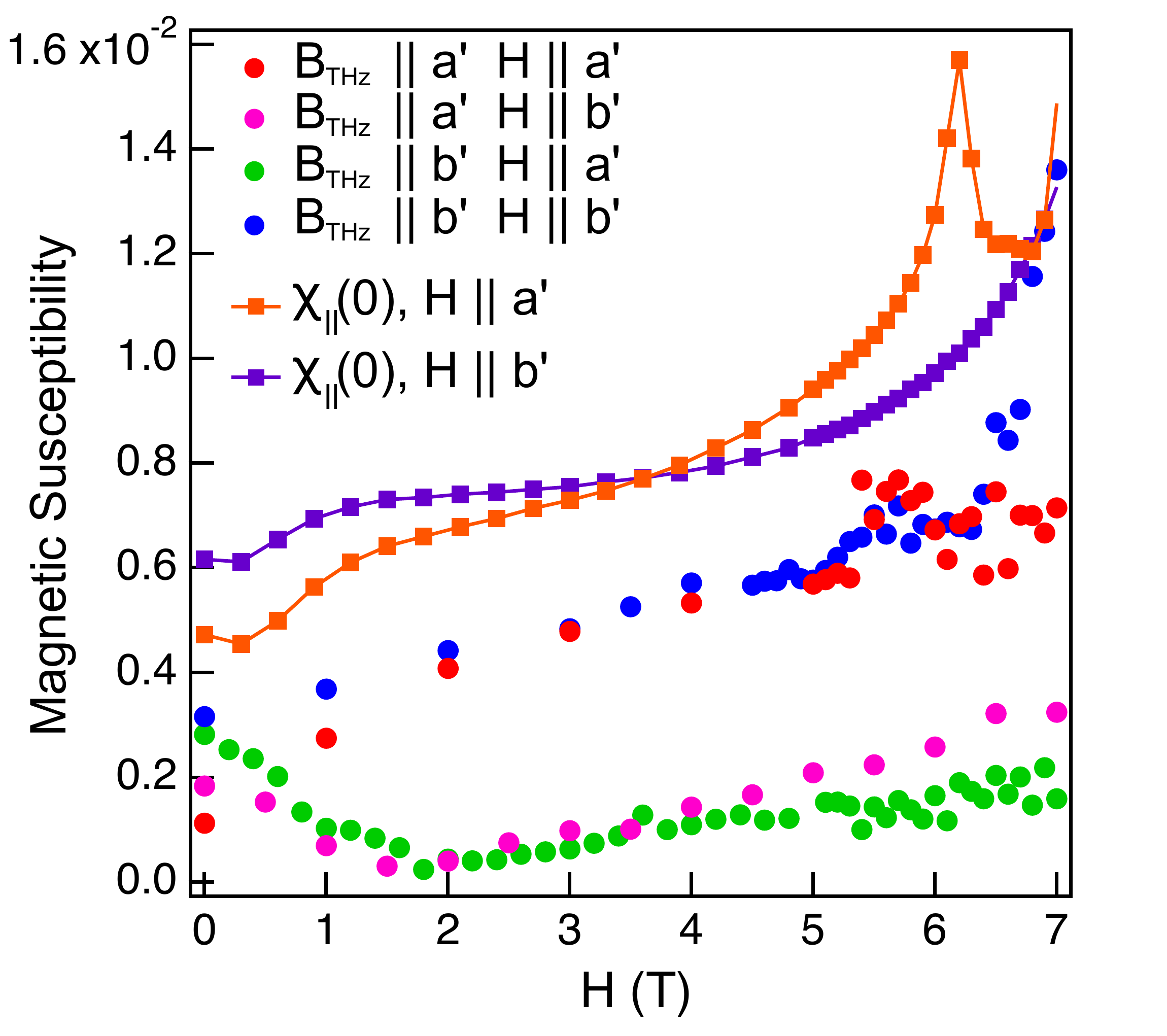}
\caption{Colored dots: Contribution of Q = 0 magnons to static magnetic susceptibility, $\chi_{sw}(0)$ as measured by fits to THz spectra for all four configurations of $\bf{H}$ and $\bf{B_{THz}}$. Orange and purple squares: Total value of $\chi(0)_{\|}$ as measured by low- frequency susceptometry for two directions.
}
\label{fig3}
\end{figure}

\par The dependence on field of the spin wave spectral weight measured with $\bf{B_{THz}}\perp\bf{H}$ is shown as well in Fig. 3, where it is seen to be strikingly different from the results for $\bf{B_{THz}}\ \| \ \bf{H}$. In this configuration the spectral weight exhibits a deep minimum at 2 T for both the $\bf a^\prime$ and $\bf b^\prime$ directions, where it nearly vanishes. The field at which this minimum occurs is the same as the field at which the crossover from the $\Omega_1$ magnon to $\Omega_2$ magnon takes place in the THz spectra. In the following section, we explain how the main features of these data can be modeled using LSWT.

\section{Theoretical Description}
\subsection{Linear spin wave theory}
The starting point for the LSWT calculations is the effective spin Hamiltonian,

\small
\begin{equation}
\begin{aligned}
H_S = &\sum_{<ij>} \left[J_1 {\bf S}_i\cdot {\bf S}_j + \Gamma (S_i^{\alpha_{ij}}S_j^{\beta_{ij}}+S_i^{\beta_{ij}}S_j^{\alpha_{ij}})+K S_i^{\gamma_{ij}}S_j^{\gamma_{ij}}\right]\\
 &+\sum_{<ij>_3} J_3 {\bf S}_i\cdot {\bf S}_j-\mu_B g \sum_{i} {\bf H}\cdot {\bf S}_i
\end{aligned}
\end{equation}
\label{eq:H}
where $\left<ij\right>$ and $\left<ij\right>_3$ denote summation over nearest neighbor and third neighbor bonds, respectively~\cite{rau2014generic, winter2016challenges, yadav2016kitaev, wang2017theoretical, winter2017breakdown}. $K$ is the Kitaev interaction, $\Gamma$ is the symmetric off-diagonal term and $J$, $J_3$ are the nearest-neighbor and third neighbor Heisenberg couplings, respectively. The $\gamma_{ij}$ are bond labels ($x$, $y$, or $z$) as shown in Fig. 1 (a) and $\alpha_{ij}, \beta_{ij}$ are the two remaining directions for each bond. Note that the magnetic field is expressed in spin-space components, for example, $\bf{H}\ \|\ \bf{a}$ is expressed as ${\bf H}=H\,(1,1,-2)/\sqrt{6}$ and $\bf{H}\ \| \ \bf{b}$ is ${\bf H}=H\,(1,-1,0)/\sqrt{2}$.

The parameters in Eq. 3 lead to a classical ground state with the observed zigzag antiferromagnetic order. We obtain the collective modes by expanding the Hamiltonian to quadratic order in the fluctuations about the ordered magnetic moment~\cite{holstein_primakoff_1940,toth_lake_2015,colpa_1978}. The spin wave theory is reliable when quantum (or thermal) fluctuations are small compared to the ordered moment, in which case the normal modes are non-interacting magnons. We obtain the theoretical THz absorption by computing the linear response of the magnons to an oscillating magnetic field (see Appendix B, section 1).

In the zigzag state, the unit cell of the honeycomb is enlarged to include four sites; as such there are four independent dispersing magnon modes. Of these, only two contribute to THz absorption, corresponding to the $\Omega_1$ and $\Omega_2$ modes discussed above. The two higher energy modes cannot be excited by the uniform in-plane THz field. This selection rule is exact, and is a result of a $Z_2$ symmetry of the zigzag state, whereby two pairs of spins within the unit cell may be exchanged (see Appendix B, section 3). Thus we do not associate the observed peaks at $L_3$ and $L_4$ with these modes.

To find appropriate values for the parameters in Eq. \ref{eq:H}, we began with the representative values chosen by Winter et al.\cite{winter2018probing, winter2017breakdown} to model INS data, and adjusted them to fit the energies of the modes seen by TDTS. We note that the parameters suggested by Ran et al.~\cite{ran2017spin}, obtained by fitting exclusively to INS spectra at the M point, yield spin wave energies at $\bf{Q}=0$ much larger than found experimentally. A linear spin wave calculation with the parameters of Winter et al. leads to an accidental degeneracy of modes $\Omega_1$ and $\Omega_2$. Refinement of these parameters is needed in order to account for our observation that these modes are split by 0.7 meV at $H$ = 0. In particular we find that fitting the spectra is accomplished by increasing the relative strength of the $\Gamma$ term, such that $\Gamma/K \sim-1$ instead of $\Gamma/K=-1/2$. A representative fit to the energies of modes $\Omega_1$ and $\Omega_2$ as a function of $H$ using the parameter set ($J$, $K$, $\Gamma$, $J_3$) = ( -0.35, -2.8, 2.4, 0.34) meV is shown as dots in Fig. 2 (f). We assume the same in-plane g-factor of 2.3 as used by Winter et al.~\cite{winter2018probing, winter2017breakdown}.

The calculated energies of the magnon modes are an excellent fit to the measured energies. Nevertheless the parameters we have chosen should not be viewed as a definitive set representing microscopic interactions. As we show below, there are sizable quantum corrections to spin wave theory, which should be viewed as based on renormalized parameters. Such renormalized interactions may be dependent on magnetic field and the wave vector of the mode. In this context the main role of the LSWT analysis is to explain the origin of defining features of the spectra, such as spectral weight ratios, zero-field splittings, polarization selection rules, and trends with increasing applied magnetic field.

\subsection{Low-field crossover}

\begin{figure*}[htp]
\centering
\includegraphics[width=1.2\columnwidth]{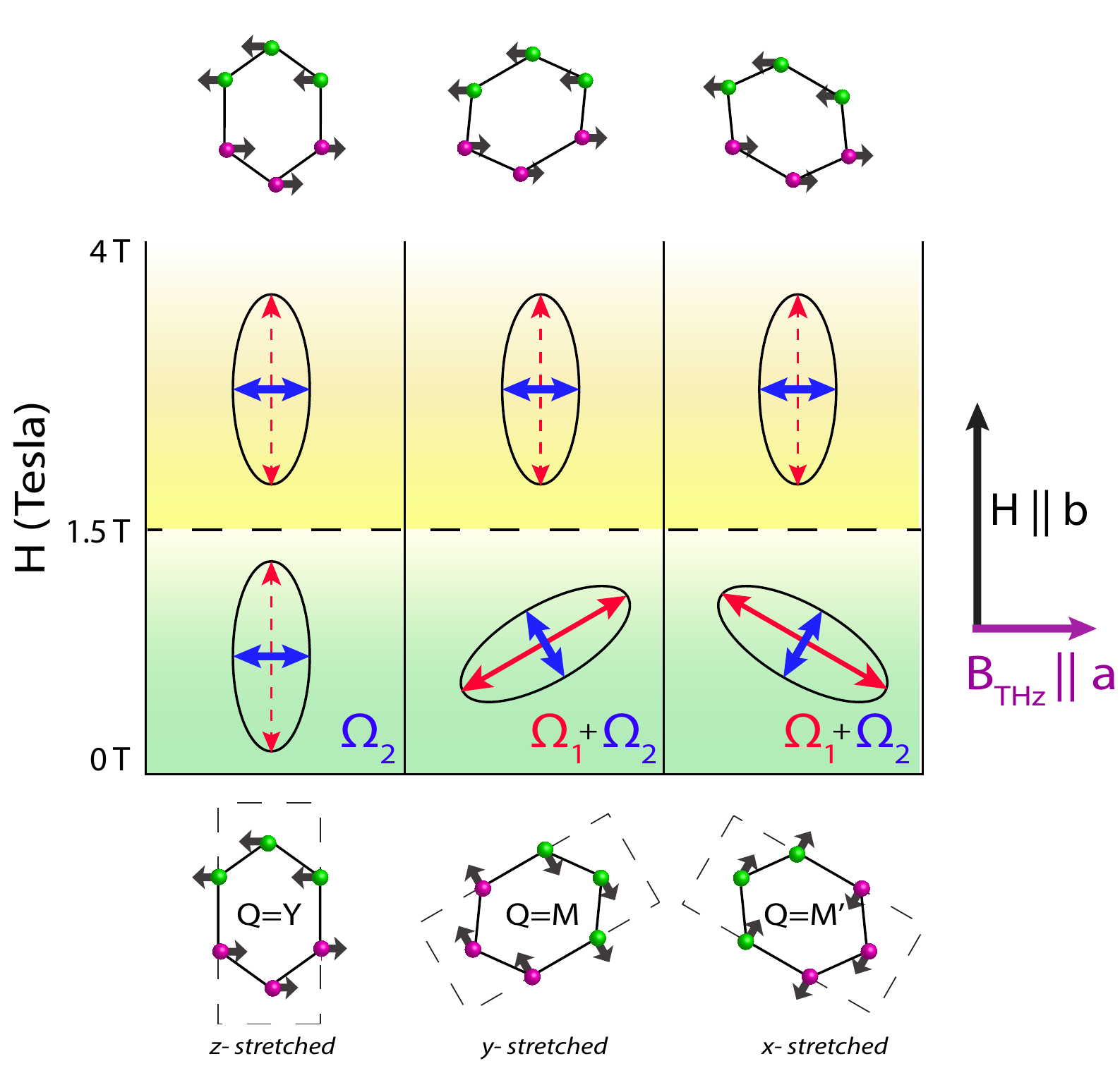}
\caption{Illustration of the evolution of the three possible zigzag states and active modes for perpendicular case ${\bf H}\ \|\ \bf{b}$, ${\bf B_{THz}}\ \|\ \bf{a}$, where $\bf{a}$, $\bf{b}$ are axes of the z-stretched domain. Bottom row of honeycombs shows preferred spin orientations at H = 0 T, with ordering wave vectors defined with respect to the z-stretched domain. The ellipses show the projection of polarization of $\Omega_1$ (red) and $\Omega_2$ (blue) onto the $ac$ plane for each domain above and below $H_X$ = 1.5 T. Solid arrows indicate a mode that absorbs for $\bf{B_{THz}}\ \|\ \bf{a}$, dashed arrows indicate a mode that does not absorb. Upper row of honeycombs shows reorientated spins above $H_X$.}
\label{fig4}
\end{figure*}
\par In the following we show that the polarization selection rules predicted by LSWT account for the intricate mode-switching behavior observed at intermediate magnetic fields, shown in Fig. 2 (a-d). The crossover at H =1.5 T coincides with the disappearance of magnetic Bragg peaks corresponding to one of the three possible orientations of zigzag order on the honeycomb lattice~\cite{sears2017phase, banerjee2018excitations}. Previously, this effect was interpreted assuming that three degenerate zigzag orientations are present as domains~\cite{sears2017phase}. Within this picture, application of a magnetic field lifts the 3-fold degeneracy, driving energetically favored domains to grow at the expense of others. The possibility that the disappearance of magnetic Bragg peaks is related to a gradual reorientation of the ordered moments within domains was also discussed~\cite{banerjee2018excitations}.

We find that a picture of gradual domain growth~\cite{sears2017phase} or spin reorientation~\cite{banerjee2018excitations} is incompatible with the abrupt changes in the THz spectra that are observed when the applied magnetic field reaches 1.5 T. Instead, our explanation of the sudden changes at 1.5 T is based on the fact that in $\alpha$-RuCl$_3$ the $C_3$ symmetry of the honeycomb lattice is broken, which removes the degeneracy of the three different possible orientations of the zig-zag magnetic order. The dependence of the relative energy of the three orientations on $\bf{H}$ will lead to a field-induced level crossing in which the wavevector of the zig-zag order will abruptly switch. In the following, we refer to this phenomenon as a ``$\bf{Q}$-flop" transition to distinguish it from the conventional spin-flop in which the spin direction changes but not the ordering wavevector.  We believe that a $\bf{Q}$-flop transition is required to account for the the abrupt changes in the THz spectra and the vanishing of certain elastic neutron peaks near 1.5 T~\cite{sears2017phase, banerjee2018excitations}. Below, we discuss in detail how the $\bf{Q}$-flop picture accounts for the unusual evolution of mode frequencies and spectral weights as a function of magnetic fields.

As mentioned previously, the breaking of $C_3$ occurs with a relatively small elongation of one of the three bond directions. We incorporate this distortion into the spin Hamiltonian by reducing the coupling constants $J$, $K$ and $\Gamma$ for the ``stretched" bond.  Breaking $C_3$ symmetry in this manner lifts the degeneracy between the three possible zigzag wave vectors, $\bf{Q}$; the zigzag with $\bf{Q}$ parallel to the direction of its stretched bond (local monoclinic $\bf{b}$ axis) is energetically favored, the two other orientations of $\bf{Q}$ related by $\pm120^\circ$ rotation are degenerate and higher in energy. This zero-field splitting plays a key role in shaping the field dependence of the THz spectra.

Our scenario for the evolution of the spectra with magnetic field is illustrated in Fig. 4, which presents a table of the energetically preferred states and active modes for each domain, for values of $H$ below and above 1.5 T. We label each bond direction by $x$, $y$, or $z$, depending on the orientation of its Kitaev interaction. The hexagons with $x$, $y$, and $z$-stretched bonds shown in the bottom row of the table illustrate the spin order of the three domains at $H = 0$, where the spins are projected onto the $ab$ plane. Our calculations show that application of a magnetic field favors zigzag orientations for which $|\bf{Q}\cdot\bf{H}|$ is largest.  At a crossover field, $H_X$, the $\bf{Q}\cdot\bf{H}$ energy gain exceeds the zero-field splitting. For $H>H_X$ the zigzag wave vector in all domains aligns with the direction selected by the magnetic field, while structural domains remain intact. The field-induced crossover is illustrated in Fig. 4 for the case where the applied magnetic field favors the domain shown in the left-hand column, in which the $z$ bonds are stretched.  For $H>H_X$ the zigzag wave vector of the $y$ and $x$ domains will reorient to the $\bf{Q}$ of the $z$-stretched domain. This process is analogous to the usual spin-flop transition in antiferromagnets, with the distinction that here the rotation involves both the direction of the moments and wave vector of the magnetic order.

The $\bf{Q}$-flop crossover described above accounts naturally for the complex evolution of the THz absorption with applied field, when we take into account the polarization and relative spectral weight of $\Omega_1$ and $\Omega_2$. As illustrated by the arrows inside the ellipses in Fig. 4, for the preferred zigzag order of a z-stretched domain (${\bf Q = Y}$), $\Omega_1$ is excited by $\bf{B_{THz}}\ \| \ \bf{b}$ and $\Omega_2$ by $\bf{B_{THz}}\ \| \ \bf{a}$. The polarization of these modes reflects an approximate symmetry with respect to exchange of \textit{x} and \textit{y} spin coordinates within the zigzag state. This symmetry is exact at zero field, and is explained in further detail in Appendix B, section 3. Furthermore, our LSWT calculations predict that the spectral weight of $\Omega_1$ is approximately a factor of six larger than that of $\Omega_2$ (as indicated by the eccentricity of the ellipses). Thus, LSWT predicts strong optical anisotropy for a single structural domain. The fact that the measured THz absorption is nearly isotropic in plane follows from the presence of the three structural domains with comparable, though unequal, population.

The state of the system for $\bf H\ <\ H_X$ is indicated by the lower row of ellipses in Fig. 4. In this regime, for all directions of $\bf{B_{THz}}$ the spectrum is dominated by the strong $\Omega_1$ mode at  2.6 meV, although $\Omega_2$ at 3.3 meV appears faintly as well. The upper row of ellipses shows the reorientation of the polarization that accompanies the $\bf{Q}$-flop crossover at $H_X$. With all the ellipses now aligned with the applied field, there is suddenly a strong dependence on the relative orientation of $\bf{B_{THz}}$ and $\bf{H}$; $\bf{B_{THz}}\  \|\ \bf{H}$ couples only to $\Omega_1$ while $\bf{B_{THz}} \perp \bf{H}$ couples only to $\Omega_2$. This results in the mode-switching from $\Omega_1$ to $\Omega_2$ that is observed only in the $\bf{B_{THz}}\perp\bf{H}$ channel. Figs. 2(e) and (f) show the evolution of the THz absorption spectra calculated with LSWT on the basis of the above model, which accurately reproduces the complex field and polarization dependent features of the experimental data.

In Fig. 5, we show that the multi-domain LSWT theory described above captures the curious deep minimum in spectral weight for $\bf{B_{THz}}\perp\bf{H}$ at 1.5 T (expressed as $\chi_\perp(0)$). The upper theoretical curve is the classical result, while the lower curve includes zero-point fluctuations of the spin 1/2 moments. The sudden reduction in spectral weight for $\bf{B_{THz}}\perp\bf{H}$ occurs when the applied field aligns the $\textbf{Q}$ of each domain, such that at $H= H_X$, $\bf{B_{THz}}$ couples only to the weaker $\Omega_2$ mode. Although the crossover predicted by the theory is sharp when compared with experiment, broadening of the $\bf{Q}$-flop crossover is expected in the presence of structural disorder. We note that our scenario is consistent with the increase of the M point spin-wave intensity at 2 T observed in INS measurements~\cite{banerjee2018excitations}.

\begin{figure}[htp]
\centering

\includegraphics[width=0.8\columnwidth]{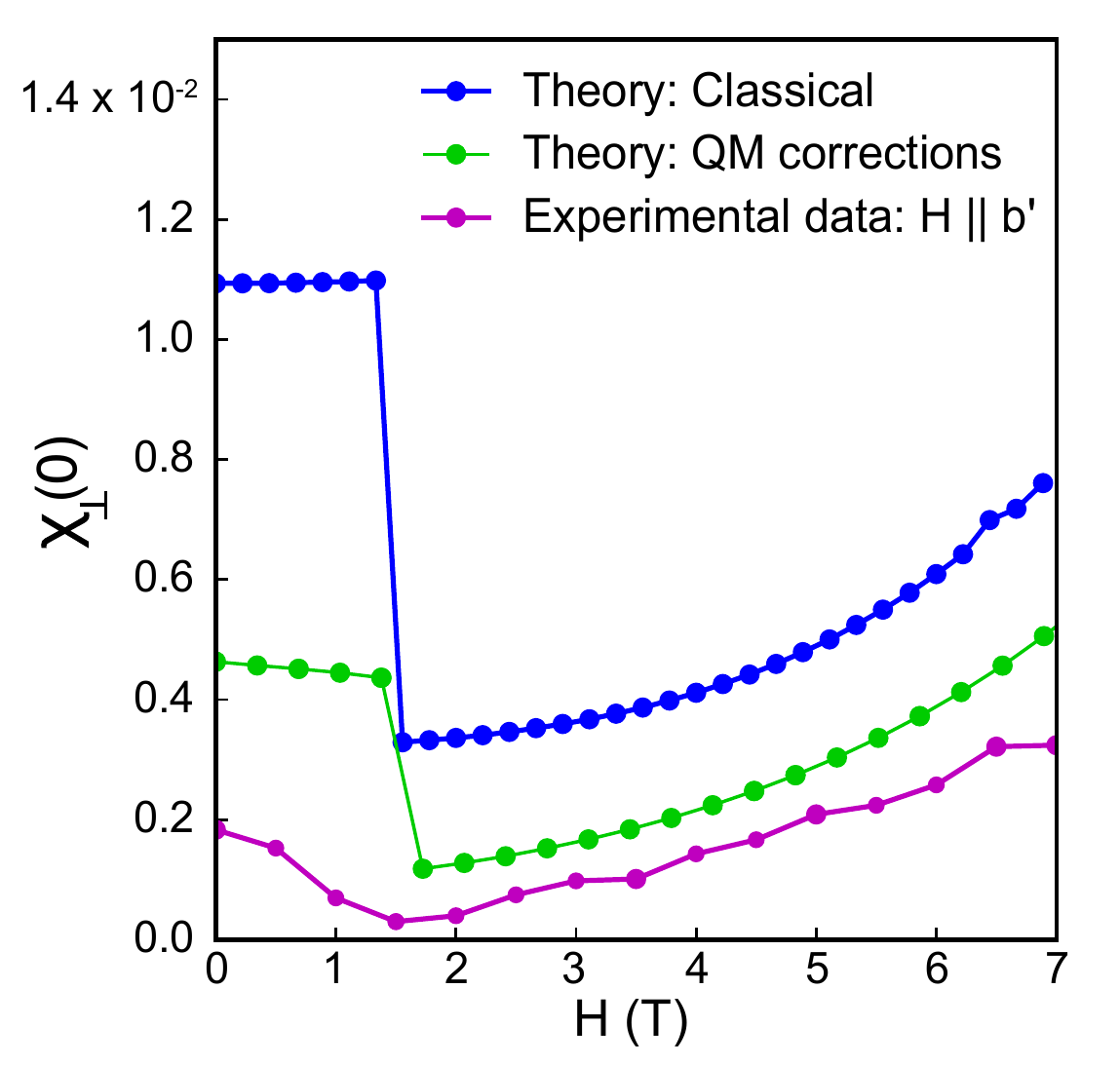}
\caption{Experimental and theoretical $\chi(0)$ demonstrating selection of the z-bond stretched ($\bf{Q= Y}$ wave vector) order at crossover field of 1.5 T for the $\bf{H}\ \| \ \bf{b}\perp \bf{B_{THz}}$ configuration. Blue: Susceptibility of the classical spin configuration. Green: Calculation of susceptibility with corrections. Magenta: Experimental values.}

\label{fig5}
\end{figure}

\subsection{Two-magnon contribution}
\par Finally, we discuss the features $L_3$ and $L_4$ that are observed in the $\bf{B_{THz}}\perp\bf{H}$ channel in the photon energy range $\sim 4-6$ meV (Figs. 2 (b) and (d)). These modes cannot be identified as single magnon excitations because of the exact $Z_2$ symmetry discussed above. However, LSWT predicts absorption by a continuum of two-magnon states in precisely this energy range (Fig. 2(f)). A further prediction is that the two-magnon absorption takes place selectively for $\bf{B_{THz}}$ parallel to the ordered moment. As shown in Fig. 4, for $H > H_X$ the moments have flopped to an orientation that is nearly perpendicular to $\bf{H}$. Thus the two-magnon interpretation of $L_3$ and $L_4$ is consistent with the selection rule seen in the data, as these features appear for $\bf{B_{THz}} \perp\ \bf{H}$ and are unobservable for $\bf{B_{THz}}\ \|\ \bf{H}$.

\par Although the selection rules show unambiguously that $L_3$ and $L_4$ are two-magnon excitations, the details of the calculated field dependence (Fig. 2(f)) differ from the data.  This is in contrast to the excellent agreement in the case of the single-magnon modes $\Omega_1$ and $\Omega_2$.  The most likely origin of this discrepancy is that while the single magnon modes are measured at $\bf{Q}$ = 0 the two-magnon absorption depends on the spin wave dispersion over the entire Brillouin zone. While our LSWT parameters reproduce the local minima at the M-points seen by INS, they do not reproduce the local minimum observed also at the $\Gamma$-point~\cite{banerjee2018excitations} (see Appendix B, section 5). Indeed, all the theoretical models of this system studied to date do not reproduce this feature of the INS data~\cite{winter2018probing, suzuki2018effective} However we find that a $\Gamma$-point minimum appears within LSWT when further interactions are added, for example second nearest-neighbor ferromagnetic coupling. Finding a spin Hamiltonian that describes all aspects of the single-magnon, two-magnon, and INS spectra is a goal for future research.

\section{Summary and conclusion}
In summary, we used polarized time-domain THz spectroscopy to track the frequencies and spectral weights of optically accessible magnetic excitations in $\alpha$-RuCl$_3$ approaching the $\sim$7.5 T transition to a spin disordered state.  The THz spectra were determined for parallel and perpendicular orientation of the static and THz magnetic fields.  We observed two sharp resonances at 2.5 and 3.2 meV and broader features in the range 4-6 meV that appear only at  applied fields of above approximately 4 T.  In the theoretical section of the paper, we showed that linear spin wave theory can account for the totality of the data, \textit{i.e}, the field dependence of spectral weights, mode frequencies, and polarization selection rules. The two lower frequency peaks are attributed to zero-wavevector magnons and the higher energy features that appear at approximately 4 T are consistent with a continuum of two-magnon excitations.

\par In our analysis, we focused on the unusual field dependence observed with $\bf{H}$ perpendicular to $\bf{B_{THz}}$, where an apparent jump in spin wave frequency from 2.5 to 3.2 meV and a deep, narrow minimum in spectral weight occur at an applied field of 1.5 T.  We showed these phenomena arise from the combination of two factors. First the $C_3$ symmetry of a perfect honeycomb is broken in the $\alpha$-RuCl$_3$ lattice, which gives rise to the presence of three structural domains. Second, the frequencies of the two optically active spin waves are split even in zero applied magnetic field;  the degeneracy of these modes seen in previous spin-wave calculations \cite{winter2018probing, winter2017breakdown} is an artifact of the parameters used in those models.  Based on these factors, we conclude that the apparent jump in frequency and spectral weight minimum arise from a $\bf{Q}$-flop crossover at 1.5 T, where the external field overcomes the anisotropy of the crystal to select a preferred ordering wave vector of the zigzag state. Although the mode jump was previously associated with Dzayaloshinskii-Moriya (DM) interaction~\cite{ponomaryov2017unconventional}, or to a sudden splitting of  modes caused by the applied magnetic field~\cite{shi2018field}, we believe that our model based on zero-field splitting and field-induced ground state energy crossing is uniquely able to account for the totality of the data. The constraints on the effective spin Hamiltonian parameters that emerge from our analysis will aid in understanding the phase diagram of $\alpha$-RuCl$_3$ and potential for existence of spin liquid ground states in this fascinating compound.

\section{Acknowledgements} We thank N. Tamura and C. V. Stan for support at the Advanced Light Source and E. Angelino for help processing the Laue microdiffraction data. We thank T. Scaffidi for useful discussions. Terahertz spectroscopy was performed at Lawrence Berkeley National Laboratory under the Spin Physics program (KC2206) supported by the US DOE, Office of Science, Office of Basic Energy science, Materials Sciences and Engineering Division under Contract No. DE-AC02-05-CH11231. Laue microdiffraction measurements were carried out at beam line 12.3.2 at the Advanced Light Source, which is a Department of Energy User Facility under Contract No. DE-AC02-05CH11231. Device fabrication and dc conductivity measurement were done at Stanford University under the Spin Physics program supported by the US DOE, Office of Science, Office of Basic Energy science, Materials Sciences and Engineering Division under Contract No. DE-AC02-76SF00515. A.L. and L.W. were supported by the Gordon and Betty Moore Foundation's EPiQs Initiative through the Grant No. GBMF4537 to J.O. at U.C. Berkeley. The work at ORNL was supported by the US DOE, Office of Science, Basic Energy Sciences, Materials Sciences and Engineering Division (J.Q.Y. and C.B.), and Division of Scientific User Facilities (A.B. and S.E.N.) under contract number DE-AC05-00OR22725. P. L. K. and D. M. acknowledge support from Gordon and Betty Moore Foundation's EPiQS Initiative through Grant GBMF44. E.A. acknowledges support from the ERC synergy grant UQUAM. D.B.'s participation in this research was facilitated in part by a National Physical Science Consortium Fellowship and by stipend support from the National Institute of Standards and Technology.
\par L.W., A.L. and E.E.A contributed equally to this work.

\appendix
\section{Experimental Details}
\subsection{Crystal Synthesis}
The sample studied was synthesized at Oak Ridge National Lab. Commercial-RuCl$_3$ powder was purified to a mixture of $\alpha$-RuCl$_3$ and $\beta$-RuCl$_3$, and converted to 99.9 $\%$ phase-pure $\alpha$-RuCl$_3$ by annealing at 500$^{\circ}$C. Single crystals of $\alpha$-RuCl$_3$ were grown using vapor transport at high temperature. The sample used for THz is roughly 5 mm x 8 mm in size and $\sim$1 mm thick. The sample exhibits a single phase transition at 7 K.
\subsection{TH\lowercase{z} generation and polarization}
THz spectroscopy measurements were performed at Lawrence Berkeley National Lab in a 7 T Janis Instruments magneto-optical cryostat. THz pulses were generated focusing an 780 nm ultrafast laser pulse onto an Auston switch, consisting of a dipolar electrode antenna patterned onto a semiconductor. An AC bias voltage is applied across the electrodes while the laser pulse excites free carriers in the semiconductor. The carriers are accelerated by the bias voltage, emitting THz radiation. The THz pulses are focused onto the sample by off-axis parabolic mirrors and the transmitted radiation is collected by a receiver antenna. The THz focus spot size is large, $\sim$5 mm, and as such only samples with large lateral dimension are suitable. To select the direction of the THz magnetic field, B$_{THz}$, the antenna polarization is fixed parallel to either the $\bf{a^\prime}$ or $\bf{b^\prime}$ axes of minimum THz transmission, as described in the main text. A grid-patterned polyethylene polarizer parallel to the antenna is placed directly before the sample (but outside of the cryostat). The DC magnetic field is applied in the $ab$ plane of the sample.
\par The index of refraction (n = 2.5 in the THz regime) and registration of optical and crystal axes were found using the methods described in the supplementary material for our previous paper, Ref. [18], which also contains a detailed description of the data analysis and of how the THz absorption is related to imaginary part of the magnetic susceptibility.

\subsection{Scanning X-ray Micro-diffraction}

\begin{figure}[htp]
	\centering
        \includegraphics[width=0.7\linewidth]{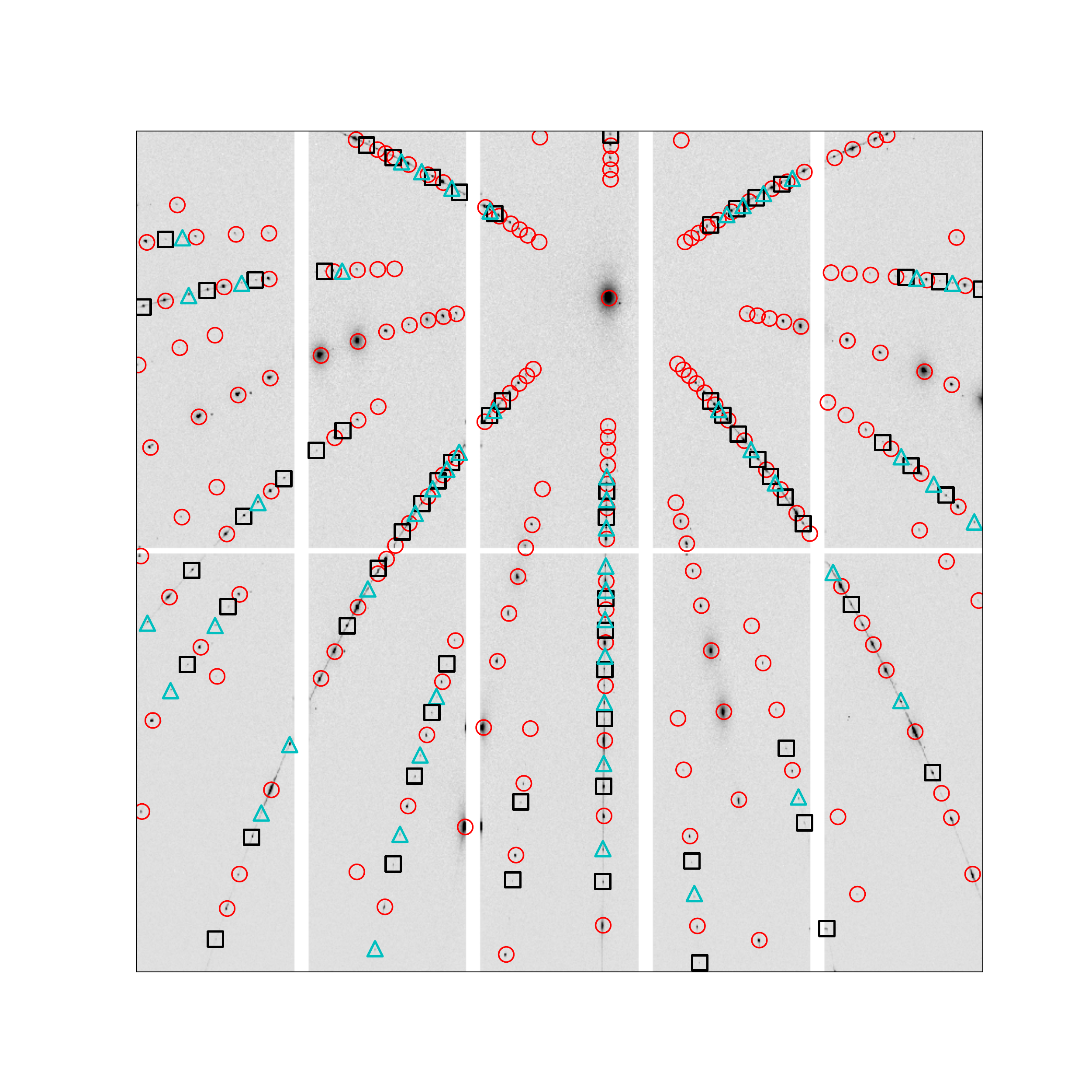}

\caption{Example Laue diffraction pattern illustrating the multi-domain character of the sample.  Red circles mark peaks successfully indexed with the dominant monoclinic domain; black squares and cyan triangles mark peaks that are unique to the monoclinic domains rotated by~$2\pi / 3$ and~$4\pi / 3$, respectively, about the sample normal with respect to the dominant domain.}
\end{figure}
In order to confirm the multi-domain character of our sample, we performed scanning X-ray micro-diffraction\cite{tamura2003scanning} on a 6 x 5 mm area of the same sample studied by THz. Using the Advanced Light Source beamline 12.3.2, a full Laue (i.e., polychromatic) diffraction pattern is collected at spots of 2~$\mu$m diameter in a 6$\times$~5~mm region of the sample. We index using the lattice parameters in the C2/m space group given in ~\cite{johnson2015monoclinic}. A representative Laue pattern from a single point is shown in Fig. S3. The diffraction peaks corresponding to three monoclinic structural domains, oriented at 120~degree intervals with respect to the sample normal, are indicated in the figure.
We find that the sample is multi-domain across the region of study.

\subsection{Measurement of $dc$ resistivity}

We measure the dc resistivity of a thin flake from the same growth as the sample measured in the main text down to T=100 K. This measurement was done at Stanford University by exfoliating flakes a few mm in size and of about 100 $\mu$m in thickness. Samples were contacted using EpoTek H20E epoxy and gold wirebonding wire. The samples were mounted on insulating sapphire substrate in a ceramic chip carrier (Kyocera PB-44567).Two-terminal DC transport measurements at low temperature showed insulating behavior, consistent with  as shown in Fig. S2. We extend the range of this measurement to temperatures lower than previously measured, where at 100 K, $\rho \sim 6.5 \times 10^{6}$ $\Omega \cdot$ cm.

\begin{figure}[h!]
        \centering
        \includegraphics[width=0.7\linewidth]{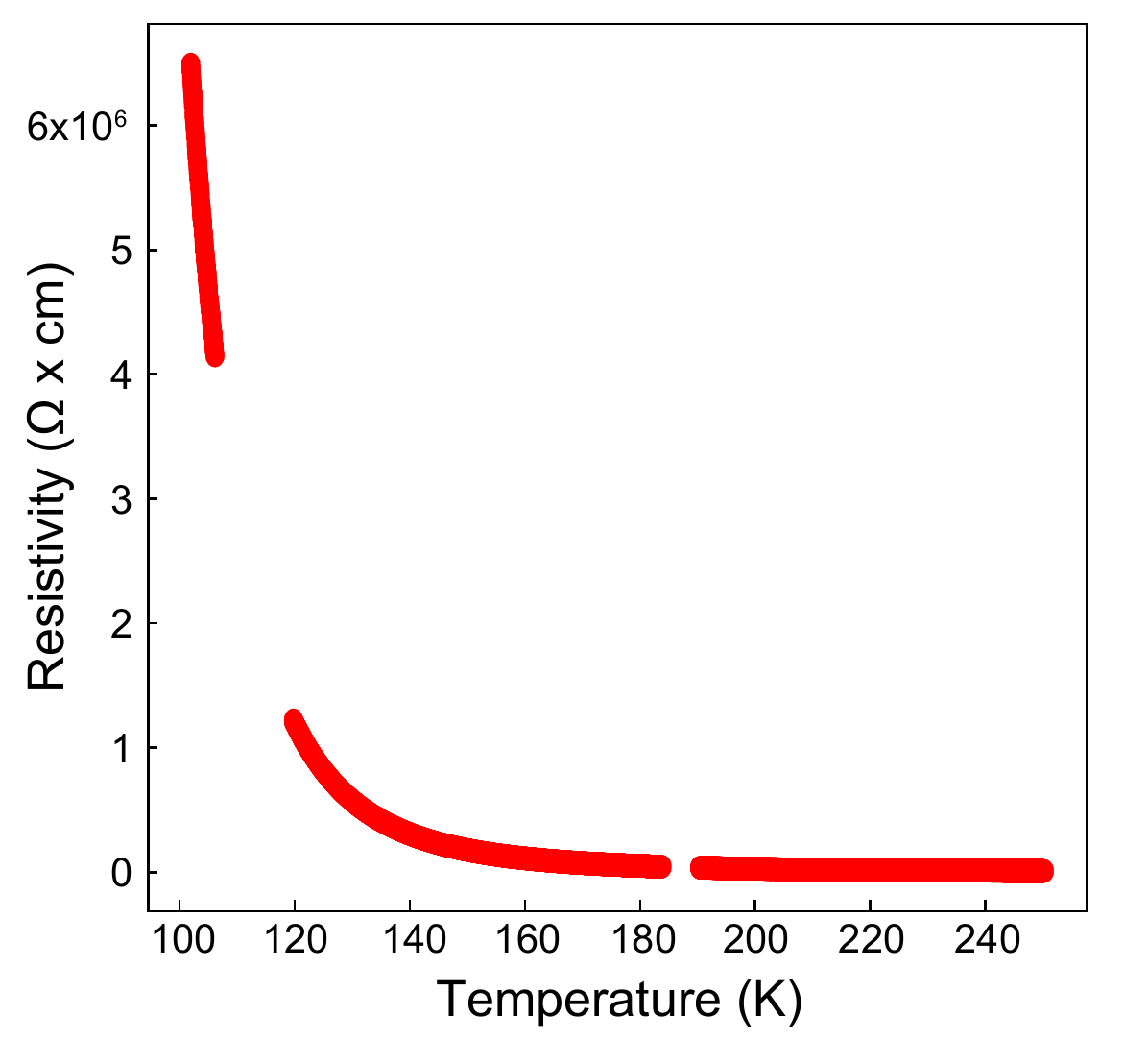}
        \caption{$dc$ resistivity as a function of temperature.}
\end{figure}

\subsection{Temperature dependence}
To clearly plot magnon contribution from the full THz absorption, as shown in the main text Figs. 1 and 2, we subtract the spectrum at T = 8 K, just above the N\'eel temperature, from the spectrum in the ordered phase at T = 4 K. This removes absorption that is temperature independent across the magnetic transition, notably the large conductivity continuum\cite{little2017antiferromagnetic, reschke2017electronic, bolens2017mechanism, bolens2018theory}. Figure S3 shows the differential spectra for just the $\Omega_1$ magnon in the parallel configuration. The 8 K - 10 K spectrum shows the change in absorption is nearly zero, and thus we select the 8 K spectrum as a reference.

\begin{figure}[htp]
	\centering
        \includegraphics[width=0.65\linewidth]{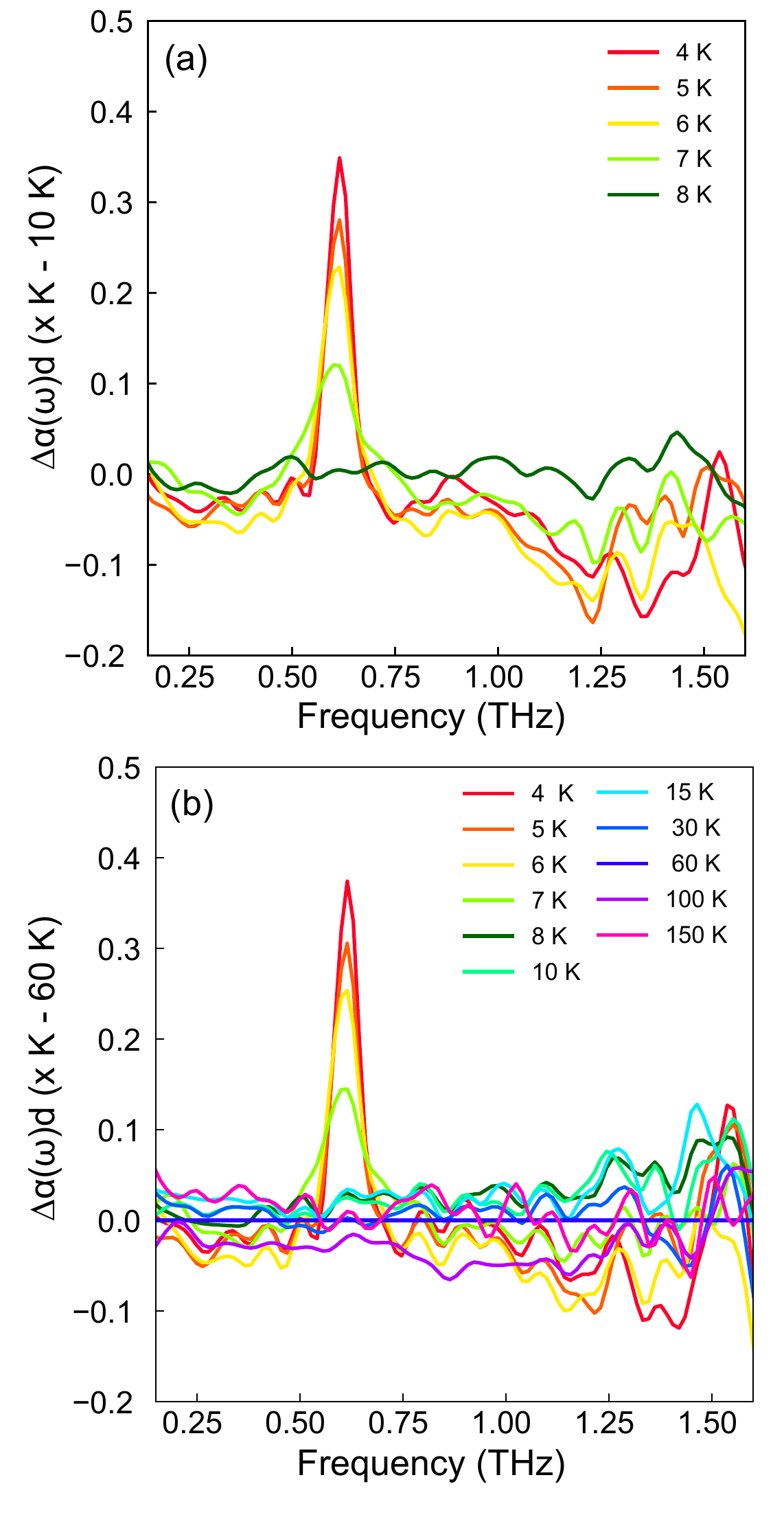}

\caption{ (a) Differential spectra showing $\Omega_1$ in the $\bf{B_{THz}\ \|\ b^\prime}$, $\bf{H} = 0$ configuration, for 4 - 8 K with the T = 10 K spectrum subtracted out. Note that for the 8 K - 10 K spectrum, the $\Omega_1$ magnon is no longer visible.  (b) Differential THz spectrum as a function of temperature for $\bf{B_{THz}\ \| \ b^\prime}$, $\bf{H} = 0$ configuration, referenced to T = 60 K. The continuum feature does not exhibit a clear temperature dependence for this sample.}
\end{figure}

We further note that for the sample studied, the temperature dependance of the conductivity continuum does not show a clear trend and is very weak. In Fig. S3, we show differential spectra, with the T = 60 K spectrum as a reference for the $\bf{B_{THz}\ \| \ b^\prime}$ up to 150 K.

Some samples have been shown to exhibit a structural phase transition in the range of 140 -160 K between monoclinic and trigonal or rhombohedral structures at low temperature~\cite{glamazda2017relation}, while some have been shown to be monoclinic at low temperature ~\cite{Ziatdinov}. Such a transition is sensitive to the stacking of the layers, and may appear hysteretic in temperature in the far infrared transmission/absporption~\cite{wang2017magnetic, reschke2017electronic}. The THz spectra of the sample used in this study do not show clear hysteresis related to the structural phase transition in this temperature range. We note that the temperature dependence of the broadband component of the absorption measured on this sample differs from that reported in Ref. \cite{wang2017magnetic}.

\subsection{Determination of magnetic susceptibility}
The total volume magnetic susceptibility, $\chi_{\|}(0)$ was measured by low-frequency susceptometry at Oak Ridge National Laboratory and was reported CGS units of emu/cm$^{3}$ for each value of magnetic field. The $\chi_{sw}^{\prime\prime}(\omega)$ measured by time-domain THz spectroscopy in SI units, where it is naturally a dimensionless quantity. To convert $\chi_{\|}(0)$ for direct comparison, we use the molar volume of $\alpha$-RuCl$_3$, 54.6 mol/cm$^{3}$ and a factor of 4 $\pi$ between the two systems of electromagnetic units.

To extract the magnon contribution to $\chi(\omega)$ from the THz spectra, we employ a the procedure described in the supplementary material for ~\cite{little2017antiferromagnetic}. In the parallel channels only $\Omega_1$ absorbs, and thus it is straightforward to subtract the raw absorption spectrum at T = 4 K and $H=$ 0 T from the spectra at H = 1 - 7 T -- removing any field-independent features. We fit the residual spectra to the subtraction of two Lorenztian lineshapes as detailed in ~\cite{little2017antiferromagnetic}. This also allows us to remove unwanted systematic features of the raw data, such as oscillations in the frequency domain which result from time-domain pulse reflections.

\par In the perpendicular orientations, the appearance of the additional peaks $\Omega_2$ and $L_3$ and $L_4$ make the field subtraction process described above unfeasible. In this case, $\chi_{sw}^{\prime\prime}(\omega)$  instead was determined by fitting a Lorentzian function to the the differential spectra across the N\'eel temperature, for example as in Fig. S3. The fits are performed at each magnetic field step, and peak parameters for each of the four absorption features are extracted.

Using these fit parameters, we evaluate the sum rule integral relating $\chi^{\prime\prime}_{sw}(\omega)$ to $\chi_{sw}(0)$ (Eq. (2) of the main text) for each absorption peak. The values for $\chi_{sw}(0)$ in main text Fig. 3 represent the total $\chi_{sw}(0)$ for all peaks, including both single-magnon and two-magnon contributions.
\begin{figure*}[htp]
        \centering
        \includegraphics[width=0.65\linewidth]{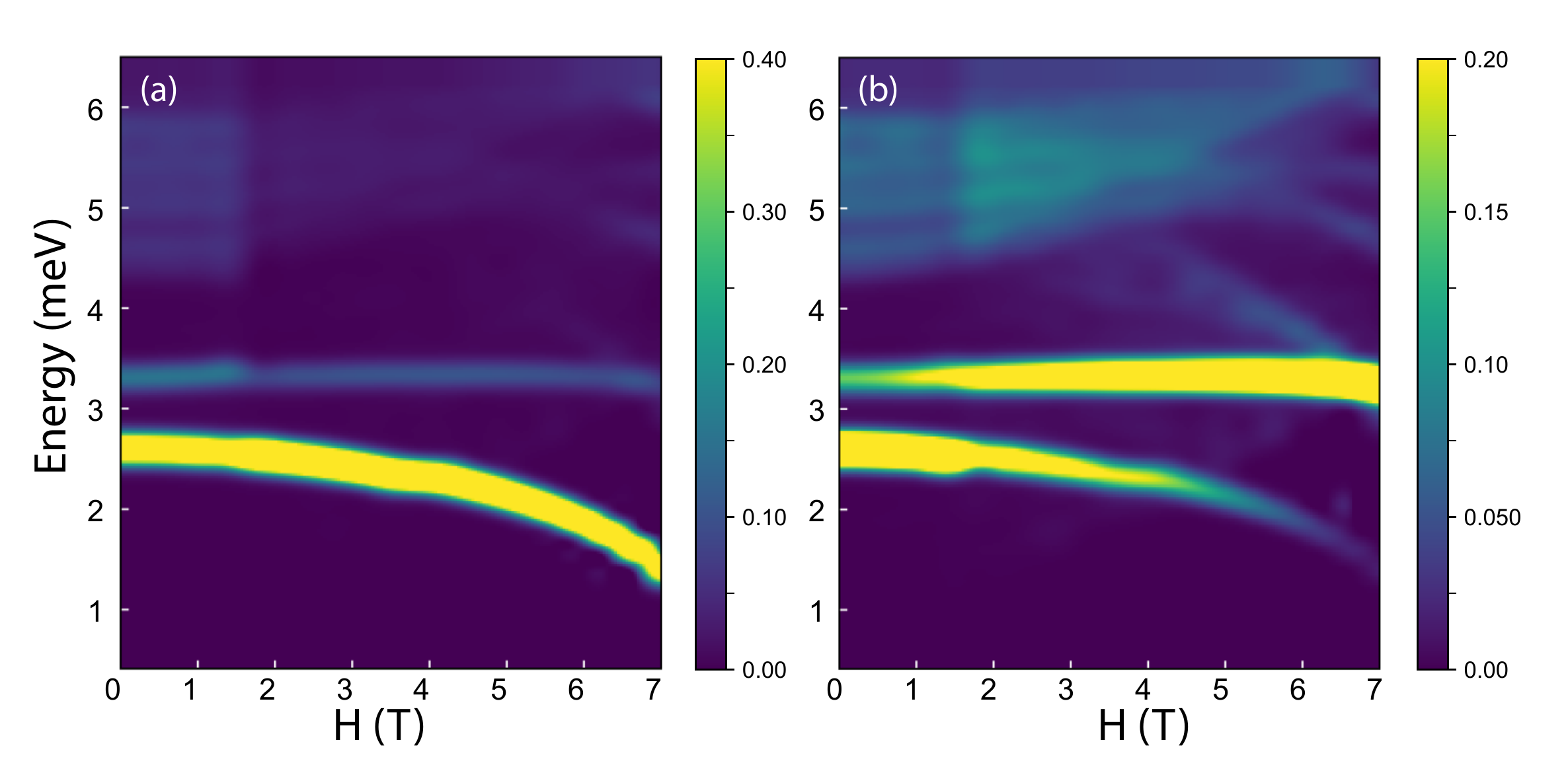}
        \caption{(a) THz absorption from LSWT for parallel channel: {\textbf H} $||$ $\bf{a}$, ${\bf B_{THz}}$  $||$ $\bf{a}$. (b) Perpendicular channel: {\textbf H} $||$ $\bf{a}$, $\bf{B_{THz}}$ $||$ $\bf{b}$. All theory plots are Gauss broadened by 0.1 meV, it is important to note that for broadening larger than 0.4 meV the mixing of polarizations in the {\textbf H} $||$ $\bf{a}$ case is difficult to observe.}
\end{figure*}
\section{Linear spin wave theory}

\subsection{Model details}
As discussed in the main text, we employ linear spin wave theory to model the antiferromagnetic resonance modes observed by THz spectroscopy. Here we discuss a few further details of this calculation. Beginning from the Hamiltonian in Eq. (3) of the main text, our spin wave approximation represents the dilute limit of magnon fluctuations above a classical spin configuration. Such an approximation is accomplished by rewriting the local longitudinal spin component as polarized minus a number operator. Consistency of the commutation relations of the spin variables and new bosonic variables fixes the rest of the dictionary:

\begin{equation}
S^z_i = S_0 - N_i=S_0 -a^{\dag}_ia_i
\end{equation}
\begin{equation}
 S^-_i-=(2S_0)^{1/2}a^{\dag}_i(1-a^{\dag}_ia_i/(2S_0))^{1/2} \approx (2S_0)^{1/2}a^{\dag}_i
\end{equation}
\begin{equation}
S^+_i=(2S_0)^{1/2}(1-a^{\dag}_ia_i/(2S_0))^{1/2}a_i \approx (2S_0)^{1/2}a_i
\end{equation}

\par Where $S_0$ is the spin magnitude and the approximation in Eq. (2) and (3) is assumed in LSWT. These redefinitions are necessarily local when the magnetic order is not ferromagnetic. This technique is detailed in the more general case of incommensurate structures by Toth et al.~\cite{toth_lake_2015}. The classical spin configuration is obtained by assuming a given zigzag order and minimizing the presented Hamiltonian. We then expand in our bosonic operators around this classical configuration. The accuracy of this approximation requires that the spin deviation remains small on the spin size.

\par Making these substitutions yields a bosonic theory with interactions, which we neglect, resulting in a quadratic theory. The quadratic Hamiltonian is not number conserving and can be solved with a Bogoliubov transformation using the technique of Colpa et al.~\cite{colpa_1978}. The zero point quantum fluctuations of the spins are captured in this approach and lead to reduction of the magnetic moment compared to its classical value. This reduction of static moment is the correction that we include when calculating the theoretical DC magnetic susceptibility (see the main text Fig. 5) and it is important in not overestimating this quantity.

\par The dynamic structure factor is given by the two spin correlation function which can be calculated with free field correlators in the bosonic language. This is related to absorption with linear response theory.
\begin{equation}
\begin{aligned}
\int_{-\infty}^{\infty}e^{i\omega t}\sum _{i,j} (<S_{i}^{\mu}(t)S_{j}^{\nu}>+
<S_{i}^{\nu}S_{j}^{\mu}(t)>)\\
 \propto \coth(\frac{\beta \hbar \omega}{2})\chi^{\prime\prime}_{\mu\nu}(\omega, Q=0)
\end{aligned}
\end{equation}
In the bosonic language this expression amounts to the evaluation of two and four point functions in a free theory. The transverse contributions are exclusively two point functions and should therefore only be sensitive to, in this approximation, the one magnon spectrum. The longitudinal component of each spin contains a number operator so evaluating the longitudinal absorption involves a four point function. A four point function in free field theory splits into an integral over pairs (in our case with net momentum zero) and necessarily generates a continuum response. It is interesting that this continuum can generate contributions that look sharp enough in width to be confused with higher energy spin wave modes or other bound states.

\par For our Hamiltonian, the 1/S corrections to spin-wave theory are small but not insignificant. They are strongest at 0-field, and close to the $\sim$7 T transition. The 0-field corrections come from spin flip occupation at momentum given by the wave vectors of the unchosen zigzag orders. The zero field corrections are due to soft fluctuation modes at the right wave-vector that would take one zigzag configuration to one of the other two degenerate states. These corrections however should not couple strongly to the individual $\bf{Q}$ = 0 modes.
\par Near the transition, the zero point fluctuation (namely, the occupation of the Holstein-Primakoff bosons $n_k=a^\dagger_k a_k$) is almost exclusively at $\bf{Q}$ = 0 and we expect quantum corrections to our calculated spectra.

\subsection{Q-flop}
As a simple model of the anisotropy present in $\alpha$-RuCl$_3$ we consider an average of LSWT results on three honeycomb patches. On each patch a different bond type is ``lengthened" by reducing all interactions across this bond by 3$\%$. This results a 0-field selection of a given zigzag order (with wave vector parallel to the stretched bond), and a competition between the magnetic field and the anisotropy of the lattice. We implement this competition by assuming the zigzag order with the minimum classical energy in each patch. This model demonstrates the physical mechanism for the abrupt change in absorption near 1.5 T as a Q-flop, a shifting of zigzag order. The average over patches is necessary for this jump to be observed for any direction of applied field. The scale of this crossover, and the fact that it occurs for multiple directions of in-plane applied field sets constraints for future studies which will necessarily consider bond-dependent couplings. Such a refined model must not select a given zigzag order too strongly, otherwise an applied field along the preferred zigzag order would yield no low field discontinuous behavior.

\subsection{Symmetries and selection rules}
We find the Hamiltonian enjoys two residual $Z_2$ symmetries even in presence of the zigzag order. The first is a sublattice symmetry. In a zigzag order the sublattice contains four spins, two on each ferromagnetic strip. The sublattice symmetry is a simultaneous switching of spins within each zigzag strip. This symmetry acts simply on the $\bf{Q}$ = 0 modes and they transform under representations of it. The higher energy modes are odd under this transformation and the lower energy modes are even. A uniform magnetic field is even under this transformation so it can only couple to the lower modes. This explains the absence of a response from the two higher spin wave modes.
\par The second selection rule is the result of a spin-space symmetry of the zigzag state. This symmetry exchanges the two spin directions not associated with the bond that joins ferromagnetic strips. For instance, if the bond that joins the ferromagnetic strips is z-type, then at zero external field there is a symmetry upon the exchange of the x and y spin coordinates. This symmetry is approximate at finite field and exact for zero field. One of the lower modes, $\Omega_1$ is odd under this transformation, while $\Omega_2$ is even. The in-plane probe field, B$_{THz}$, if applied parallel or perpendicular to the ordering wave vector will be respectively odd or even under this transformation and couple to a mode of the same parity.

\subsection{Theoretical spectra for ${\bf H\ } || {\ \bf a}$}
The theoretical absorption for ${\bf H\ \|\ b}$ is shown in Fig. 2 of the main text. Here we present the ${\bf H\ \| \ a}$ orientation, in the parallel and perpendicular channels for a multi-domain model. An important distinction between these cases is that a field along the $\bf b$ axis selects $\bf{Q}$ = Y (z-bond stretched) and is parallel to it so the selection rules apply, whereas a field along the $\bf a$ axis selects $\bf{Q}$ = M,M' (x or y bond stretched) and is neither parallel or perpendicular to these directions. We thus see some mixing of the polarization of $\Omega_1$ and $\Omega_2$; both modes are visible even after the 1.5 T crossover. Another way to visualize this is to consider the fact that zigzag orders can be rotated into each other with a joint spatial and spin rotation of 60 degrees and 120 degrees about the out of plane direction respectively. This action connects the zigzag strips which are angled 60 degrees relative to each other and appropriately changes the bond types. Therefore the spin wave modes which exclusively couple to directions $\bf a$ or $\bf b$ will be rotated 120 degrees into modes which have mixed absorption with respect to these directions. This effect weakly observed in the experimental data shown in the main text Fig. 2 (b), where $\Omega_1$ persists up to $\sim$3 T. As mentioned in the main text, the experimental results indicate that the modes broaden more rapidly for the $\bf{H\ \|\ a^\prime}$ direction than for the $\bf{H\ \|\ b^\prime}$. Classically, our model has a transition for fields along the $\bf{H\ \|\ a}$ direction at around 7.5 T and for the $\bf{H\ \|\ b}$ direction at 10 T. However the exact diagonalization results of Winter et al. \cite{winter2018probing} suggest the difference in critical field between the two directions will be smaller after accounting for quantum effects. For this reason we expect that for fields of about 7T, in the direction of $\bf{a^\prime}$, the, corrections beyond semiclassical linear results become relevant.

\subsection{Momentum dependence}

We note that the minimal LSWT model we use to describe the THz spectra at $\bf{Q}$ = 0 does not reproduce a feature present in the INS data at higher values of Q. The low field spectra in ~\cite{banerjee2018excitations} show that long the line in momentum space from the $\Gamma$ point to the M point, the lowest spin wave mode appears to increase in energy before decreasing again. This feature is not present in the standard well-studied models. The dispersion of the modes across the Brillioun zone for our LSWT model, at $\bf{H}$ = 2 T, is shown in Fig. S5, where local minima appear at the M-points but not the $\Gamma$-point.
\begin{figure}[h!]
        \centering
        \includegraphics[width=0.8\linewidth]{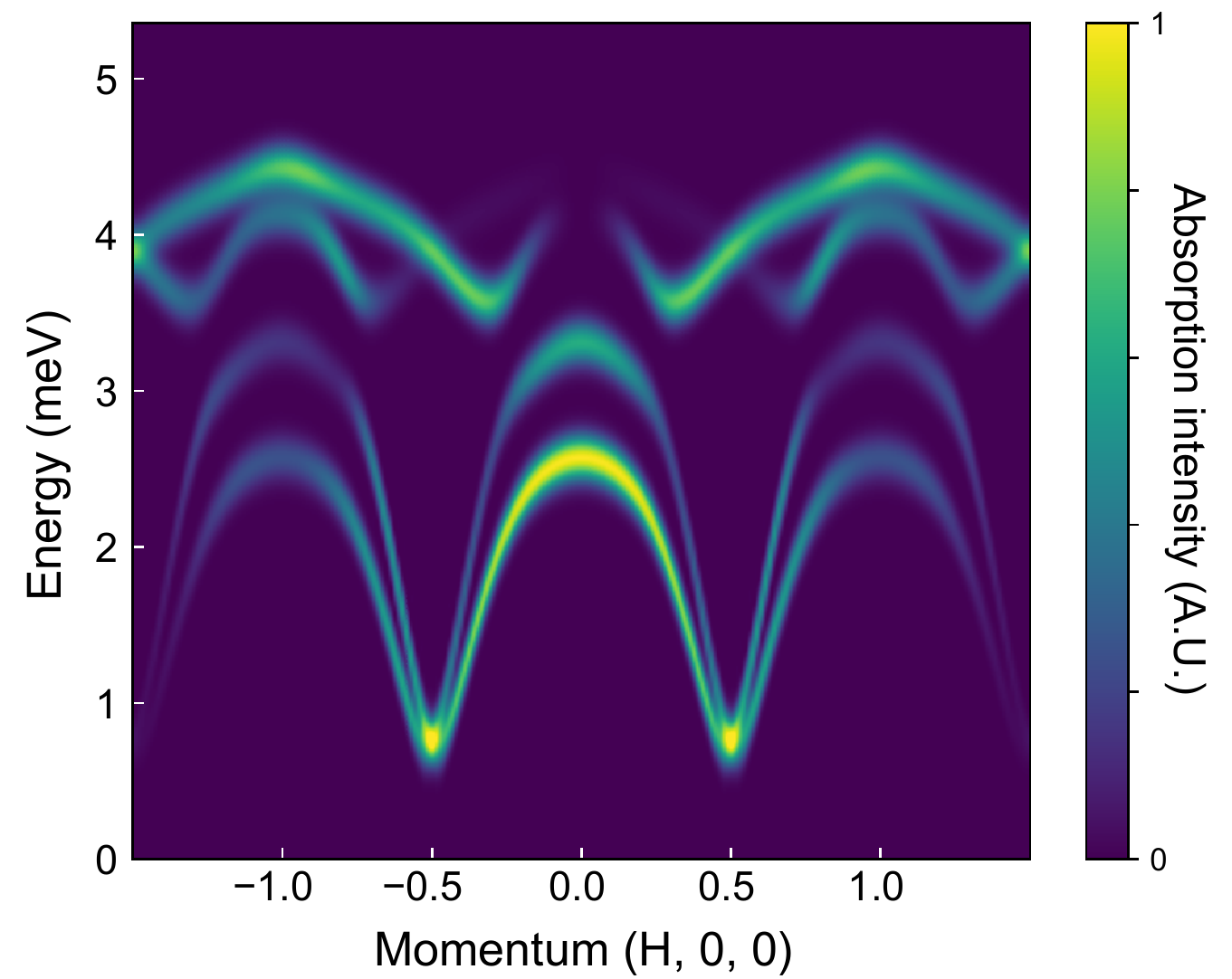}
	\caption{Theoretical dispersion at $\bf{H}$ = 2 T applied along the ${\bf b}$-axis, calculated with LSWT parameters used to model THz data in the main text. We assume the zigzag order that this field selects. }
\end{figure}
\par We find that a local minimum at the $\Gamma$-point may be introduced by adding a second nearest neighbor ferromagnetic Heisenberg term on bonds perpendicular to the order wave vector on the order of .3 mev. Anisotropic second neighbor terms of this order and larger are predicted by ab initio studies ~\cite{winter2016challenges}. Including such terms has the additional benefit of increasing the M-point gap closer to observed values. We would like to emphasize that in modeling the full dispersion, such terms shouldn't be ignored due to their small magnitude because for each site, there are six second neighbor couplings. Further study is required to fit such terms to experimental data.

\bibliography{RuCl}

\end{document}